# Second Life: Game, Simulator, or Serious Game?
## Second Life: Jogo, Simulador ou Jogo Sério?

Renato P. dos Santos[1]

*Abstract*

Ten years after being launched in 2003, it is still a matter of discussion if Second Life (SL) is a virtual world, a computer game, a shopping site or a *talker* (communication software based on a MUD). This article reports on an exploratory case study conducted to analyze the Second Life (SL) environment and determine into which of *training simulators, games*, *simulation games,* or *serious games* categories SL fits best, by using Narayanasamy et al. and Johnston and Whitehead criteria. We also examine the viability of SL as an environment for physical simulations and microworlds. It begins by discussing specific features of the SL environment relevant to its use as a support for microworlds and simulations, as well as a few differences found between SL and traditional simulators such as *Modellus*, along with their implications to simulations, as a support for subsequent analysis. We conclude that SL shows itself as a huge and sophisticated simulator of an entire Earthlike world used to simulate real life in some sense and a viable and flexible platform for microworlds and simulations.

*Resumo*

Dez anos após seu lançamento em 2003, ainda se discute se o Second Life (SL) é um mundo virtual, um jogo de computador, um site de compras ou um *talker* (programa de comunicação baseado em um MUD). Este trabalho relata um estudo de caso exploratório conduzido para analisar o ambiente Second Life (SL) e determinar em qual das categorias *simulador de treino, jogo*, *jogo de simulação* ou *jogo sério* o SL se enquadra melhor usando os critérios de Narayanasamy et al. e de Johnston e Whitehead. Também se analisará a viabilidade do SL como suporte plataforma para micromundos e simulações. Inicialmente, serão discutidas em detalhe algumas características específicas do ambiente SL relevantes para seu uso como plataforma para micromundos e simuladores, bem como algumas diferenças encontradas entre SL e simuladores tradicionais, tais como o *Modellus*, bem como suas implicações para simulações e como suporte para a análise subsequente. Conclui-se que SL apresenta-se como um enorme e sofisticado simulador de todo um mundo semelhante à Terra, utilizado por milhares de usuários para simular a vida real de algum modo e uma plataforma viável e flexível para micromundos e simulações.

**Palavras-chave:** Second Life, ensino de Física, mundos virtuais, micromundos físicos, simulações computacionais.

[1] **Renato P. dos Santos** is a Doctor of Science (Physics) and Associate Professor at PPGECIM - Programa de Pós-Graduação em Ensino de Ciências e Matemática – ULBRA. Universidade Luterana do Brasil/PPGECIM. Av. Farroupilha, 8001. 92450-900 Canoas, RS. E-mail: fisicainteressante@gmail.com.

*[I]t does not really exist. But right now, millions of people are walking up and down it …. [O]f these billion potential computer owners, maybe a quarter of them actually bother to own computers, and a quarter of these have machines that are powerful enough …. [T]hat makes for about sixty million people who can be [inworld] at any given time.* Neal Stephenson, Snow Crash (1992, pp. 23-25)

## INTRODUCTION

Neal Stephenson's science fiction novel *Snow Crash* (1992) introduced readers to the concept of the Metaverse, an online environment that was a real place to its users, one where they interacted using the real world as a metaphor and socialized, conducted business, and were entertained (ONDREJKA, 2004a).

In typical single-player video games, the game world stops when the player shuts down the computer for the night and only resumes when the player begins playing the game again the next day. Virtual worlds like *Second Life* (SL), on the contrary, are *persistent* in the sense that they continue to exist and evolve in real time around the actions of the other players who are logged in (TSENG, 2011).

Moore et al. (2008, Foreword, p. iii) suggests that new 3D worlds such as SL tend to replace the 2D Internet we know. However, Azzara (2007) argues that they represent instead "the emergence of a brand new communications paradigm". This author says that, while our present Web is based on human interaction with automated systems (such as eBay, Amazon, and Google), "virtual worlds are all about human [presential] interaction with other humans" in a 3D immersive virtual environment.

There are more than 700 (MMORG Gamelist-All Listed Games, s.d.) *game worlds*, specifically created for entertainment, e.g. *World of Warcraft*, and over 50 different Multi User Virtual Environments (MUVE) currently available (TAYLOR, 2007), created to simulate real life in some sense. SL is surely not the one with the biggest user population (TAYLOR, 2007). Nevertheless, among them, SL, followed by *OpenSim* and *Active Worlds*, stands out as the platform that offers more services and tools for developing applications with quality (REIS et al., 2011), in terms of realism, physical verisimilitude, scalability, interaction, user-friendliness, and safety. In fact, differently from other virtual worlds where physical laws are not seriously taken into account, objects created in SL are automatically controlled by the powerful Havok™ physics engine software (HAVOK.COM, 2008). As we are particularly interested in Physics simulations, these points have leaded us to choose it for our research purposes.

The ease in which new users can join SL combined with support from several educational and library groups, discussion forums and a comprehensive range of free communication, graphics, design and animation tools make many educators from around the world see SL as a versatile environment to conduct pedagogical activities (CALOGNE; HILES, 2007).

Findings suggest that SL offers flexible and wide-ranging possibilities for simulations in mechanics (BLACK, 2010). Andrew Linden, Co-Founder of Linden Lab, said that there are lots of apparent potential of SL as a medium to teach Physics, but there are many hurdles (LINDEN, 2010). Paradoxically, however, while Science has often been reported as preferential domain for virtual learning environments, most MUVEs have been used as mere places for exploration and inquiry underlining group interaction, leaving learning by interacting with the environment and modifying it in a constructivist approach as a secondary objective (VRELLIS et al., 2010).

In a rapid tour through many virtual spaces dedicated to Science, one will almost only find real world replica, mere institutional presence, with traditional classrooms, virtual boards exhibiting 2D PowerPoint™ presentations, and video seminars. We tend to agree with Kapp (2007) that having a bunch of people virtually sitting in a classroom is not the best use of Second Life or any other metaverse.

For the scientific minded visitor, a tour to *SciLands* is *de rigueur*. It is a mini-continent in SL devoted exclusively to science and technology (SCILANDS, 2007). The SciLands was created by a group of scientists and educators from universities, research laboratories, museums, and scientific agencies, such as *National Physical Laboratory* (UK), *NASA*, San Francisco's *Exploratorium* hands-on science and technology museum, *Elon University*, *The University of Denver*, and *The International Spaceflight Museum (ISM)*. According to Medeiros (2008), SciLands delivers much of the science content on SL, which ranges from seminars on nanotechnology to weekly live discussions of US National Public Radio's *Science Friday show*. Unfortunately, it received extra media coverage on February 2012 when it was disabled (deleted) by Linden Lab due to non-payment of virtual land tier fees after troubles of lack of funding and the Linden Lab refuse to grant non-profit tax exemption status to the virtual museum (AU, 2012a). Thanks to the community response in the blogosphere, however, Linden Lab agreed to put the museum back on previous terms one month later (AU, 2012b).

In SL (or in any other metaverse), the real advantage is using the platform to do innovative things that could not otherwise be done in a classroom that reach into the pupil's imaginations (FREITAS, DE; GRIFFITHS, 2009). For Doherty et al. (2006), we can use these environments to place appropriately sized avatars to move freely around and examine and interact with objects and phenomena in three-dimensional simulations of the difficult to comprehend worlds of the very large, such as planetary systems, and micro environments of cellular and even nanoscale worlds. An example is the Brownian motion[2] simulation, available at *Exploratorium* in SciLands (SCILANDS, 2007), in which the avatar may enter into one of the particles and experience its motion from inside as if he were in a *holodeck*[3] (SWARTOUT et al., 2001).

In this work, we examine the possibility of the SL environment as a support tool for physical microworlds and simulations as well as the challenges the environment presents for effective use by instructors and learners.

## *METHODOLOGY*

Dos Santos (2012) analyzed to which extent SL possesses desirable characteristics expected of any physics simulator environment. In this work, we conducted an exploratory case study (FLYVBJERG, 2011) to examine the viability of SL as an environment for physical simulations.

This work begins by discussing some characteristics of the SL environment as a support for subsequent study. We will also point out a few differences found between SL

---

[2] Brownian motion is the random movement of microscopic particles suspended in a liquid or gas, caused by collisions between these particles and the molecules of the liquid or gas, named for its identifier, Scottish botanist Robert Brown (1773-1858) (BROWNIAN motion, 2005).

[3] In the fictional *Star Trek* universe, a holodeck is a form of *holotechnology* [some kind of virtual reality based on holography] designed and used by Starfleet. (OKUDA; OKUDA, 1999, p. 193). Inspired by it, various research projects such as the *Mission Rehearsal Exercise (MRE) Project* at the USC Institute for Creative Technologies (ICT) (SWARTOUT et al., 2001) are on their way to create holodeck-like virtual reality training environments..

and traditional simulators such as *Modellus* (TEODORO et al., 1997), along with their implications to simulation of Mechanics. A few concrete examples of simulations in SL will be presented briefly in order to clarify and enrich both discussion and analysis. In addition to the author's own experiences, this work will be based in the following sources:

*Second Life Physics: Virtual, real, or surreal?* (DOS SANTOS, 2009)
Creating Your World (WEBER et al., 2007).
Havok Physics Animation v. 6.0.0 PC XS User Guide. (HAVOK.COM, 2008);
Second Life: Guidelines for educators webpage (SECOND Life: Guidelines for Educators, 2008);
*LSL Wiki.* (http://lslwiki.net/lslwiki/);
*Second Life Wiki.* (http://wiki.secondlife.com/wiki/);
*Second Life Wikia.* (http://secondlife.wikia.com/wiki/);

Afterwards, we will use Narayanasamy et al. (2006) and Johnston and Whitehead (2009) criteria to analyze the SL environment and determine into which of *training simulators, games*, *simulation games,* or *serious games* categories SL fits best.

## *FEATURES OF SL RELEVANT TO BUILDING MICROWORLDS AND SIMULATIONS*

> "*Morpheus: … yet their strength and their speed are still based on a world that is built on rules.*" (IRWIN, 2002)

On June 2013, the world of SL was made up of 6,726 uniquely named *Regions* (SHEPHERD, 2013a) - virtual parcels of land representing an area of 256 m x 256 m (65,536 m$^2$ or about 16 acres) (SIMULATOR, s.d.). These Regions are linked together to make a continuous area of about 460 km$^2$ known as the *Mainland*, almost as large as the Principality of Andorra. Including the *Private*, *Homestead* and *Openspace* regions, that figure rises to 1,742.21 km$^2$ on 27 October (SHEPHERD, 2013b), not much smaller than the country of Luxembourg.

It is interesting to remember here that 256 km was the distance between *Express Ports*, the stations of the monorail that ran the entire length of Stephenson's Metaverse *Street* (STEPHENSON, 1992). It is also worth mentioning that SL do still have *Telehubs* today even after allowing free "direct teleport" from one point to another in 2005 (HISTORY of Second Life, 2011). SL creators were clearly trying to sell it as the realization of Stephenson's vision, as per (ONDREJKA, 2004a) words: "Second Life […] is taking the first steps on the path to the Metaverse."

Each Region is rendered by a single process, which is running on Linden Lab servers (SIMULATOR, s.d.), that simulates a full rigid-body physical dynamics, including gravity, elasticity, and conservation of momentum (PHYSICS engine, 2008), and an accurate, polygon-level collision detection of all the objects in the column-space above its land.

Each server is edge-connected to up to four other machines as a grid of computers. This grid computes a simplified solution of the Navier-Stokes equations to simulate the motion of winds and clouds that time-evolve across the entire world (ROSEDALE; ONDREJKA, 2003; ONDREJKA, 2004b). In addition to rendering the land and simulating its physics and weather, this grid keeps track of all the millions of independent *primitives*

(the building blocks of SL that allow residents to create anything they imagine) (PRIMITIVE, 2010) of all the avatars within its Region (ONDREJKA, 2004b). It also caches, delivers object and texture data within the Region, and runs user's scripts and streaming routines to send back all the data needed to view the world to anyone's client who is connected (ONDREJKA, 2004b). As a result, the SL "Sun" rises and sets each 4 Earth hours (if the *Region Environment Settings* have not been altered (REGION environment settings, 2011)) always directly opposite a full Moon ("llGetSunDirection," 2009), objects fall under the effect of gravity, trees and grass blow in the wind and clouds form and drift (ONDREJKA, 2004b). Therefore, SL attempts to model the surface of an Earthlike world in a reasonably life-like way (ONDREJKA, 2004b).

Once logged through the client software, usually called a *viewer*, SL users (called *residents*) can walk around, explore the world, enjoy the 3D scenery, fly, drive cars and other vehicles, interact with other avatars, play or create objects. There is a wealth of resources for building complex objects, with many different textures, such as chairs, clothes, jewels, vehicles, guns and even entire buildings. In fact, well over 99% of the objects in SL are user created, and users have responded positively to the idea of creating the world that they live in (ONDREJKA, 2004a), which has been characterized as a shift of culture, from a media consumer culture to a participatory culture (JENKINS et al., 2006).

Differently from other virtual worlds where physical laws are not seriously taken into account, SL is possibly the most realistic virtual environment in the market given that objects are controlled by the Havok™ physics engine software (HAVOK.COM). This powerful software has been used in the creation of many internationally acclaimed films over the years such as *The Matrix*, *X-Men: First Class*, *Harry Potter and the Deathly Hallows*, and *Prometheus* (HAVOK.COM), as well as games such as *Saints Row IV™*, *The Last of Us™*, and *Assassin's Creed III®*, among others (HAVOK.COM). From Havok version 1, SL evolved to its present version 2010, which was merged into the Mesh server branch on 2011 (HAVOK 2k10 Beta Home, 2011).

A physics engine is a program that simulates Newtonian object collisions and interactions in a mathematically and computationally simulated virtual environment (PHYSICS engine, 2008). It is capable of simulating gravity, elasticity, the conservation of momentum between colliding objects, and the obedience to physical laws and principles such as gravity, buoyancy, mass, and friction, in such a way that an avatar cannot pass through walls and stones tossed into water behave as expected. Without a physics engine, an avatar in the 3D environment would simply move straight through anything in its path, as it happened in defunct *Google Lively* virtual world (GOOGLE Inc., 2008), and would fall through the ground and forever.

However, these features are not the physics engine core purpose in Second Life (PHYSICS engine, 2008). At its most basic level, it is used simply to determine empty space from filled space and to ensure that the avatar walks up and down hills on the terrain and stands upon stairways or walkways above the ground with confidence of their solidity (PHYSICS engine, 2008).

All this processing effort is divided between the simulator and the viewer. The simulator's job is to run the physics engine, detect collisions, keep track of where everything is, and send locations of content and updates to the viewer when specific changes occur. The viewer's job is to handle locations of objects, to obtain velocities and other physics information, and to do simple physics to keep track of what is moving where (LIMITS, 2012).

In recent years, according to Narayanasamy et al. (2006), the cross-boundary technology exchange between game and simulation technology along with other reasons have contributed towards the confusion as to what makes a Simulation Game and what makes a Simulator, as well as to hybrid applications such as Serious Games (ABT, 1970).

However, as Maier and Größler (2000) point out, there is still some confusion about what is meant by 'simulation' in the literature, as well as in discussions among scientists and practitioners. According to these authors, this misunderstanding, arising from a number of factors including academic backgrounds, marketing concerns and unreflective adoption of terms originally used with other intended meanings, further confounds the already complex issue of the efficacy of these objects. As an example, these authors cite Papert's (1980) understanding of microworlds.

For Papert (1980), physical microworlds are computer-based interactive learning environments where learners are allowed to play with an infinite variety of alternative laws of motion. Learners evolve along a Piagetian learning path from the historically and psychologically significant Aristotelian ideas, through the 'correct' Newton's Laws, the more complex Einstein's Relativity Theory, and even to laws of motion that students could invent for themselves, without being force-fed 'correct' theories before they are ready to invent or understand them (PAPERT, 1980). By integrating history of science, these microworlds would provide pedagogically effective experimentation with successive physical laws other than Newton's ones, in a Piagetian historical psychogenetical framework, as proposed by Papert (1980). For Papert, before being receptive to Newton's laws of motion, students should know some other laws of motion, not as complex, subtle, and counterintuitive as Newton's laws.

For Maier and Größler (2000), the term microworld should rather be used for learner-centered, modeling-oriented software packages, which are instruments to construct and simulate models.

However, for Rieber (1996), microworlds have two distinctive characteristics that may not be present in a simulation. First, a microworld presents itself as the 'simplest case' of some domain of interest, providing the learner with the means to reshape the microworld in order to explore increasingly sophisticated and complex ideas. Second, a microworld must match the learner's cognitive and affective state. In contrast, a simulation is determined by the content or domain it seeks to model and is usually judged based on its fidelity to the domain it simulates. Another important difference between microworlds and simulations is that, in the former, the student is encouraged to think about it as a 'real' world and not simply as a simulation of another world (for example, the one in which we physically move about in) (MICROWORLDS, 2011).

The literature lists various different experiences of teaching with microworlds from diSessa's (ABELSON; DISESSA, 1981) 'Dinatarts' (dynamic turtles) to Masson's historical microworlds (MASSON; LEGENDRE, 2008) and dos Santos'microworld (2013) (for a review, see e.g. (HEALY; KYNIGOS, 2009)). Unfortunately, while SL is itself a huge and sophisticated simulator of numerous real-world situations, its widely exalted 'potential' (e.g., (CONKLIN, 2007)) for simulations that promote the teaching of physics does not seem to have been understood yet – let alone be accomplished. Black (2010) is an honorable exception for having succeeded in simulating the motion of an object under the influence of simple forces while floating free of gravity.

Differently from other virtual environments, more than just allowing the construction and manipulation of objects such as fountains, weapons, or vehicles, as in other virtual environments, SL offers interactivity features to add to the objects. Users can make them move, hear, speak, change color, size or shape, and 'communicate' with other objects through its *Linden Scripting Language* (LSL)[4]. Its structure is based on Java and C, and it provides nearly four hundred functions, among which several of interest to Physics studies in this environment. For example, *llGetPos* returns the position of the object in the region (llGetPos, 2009), *llGetOmega* returns the angular velocity of the object (llGetOmega, 2010), and *llSetForce* applies a force to the object (llSetForce, 2009). They are usually broken into two categories (DYNAMICS, 2007): **Kinematic functions**, which operate on non-physical objects (see below) and generate motions without consideration of forces and torques; and **Kinetic functions**, which generate forces and torques on physical objects, (see below) resulting in its motion.

However, some important points should be taken into account when one wants to build a simulator or a microworld in SL. See also (DOS SANTOS, 2009) for a deeper discussion on these points.

*Physical objects are different*

SL computes a full rigid-body simulation of the world, but when creating, residents are free to ignore most real-world physical constraints, such as gravity and collision between objects (ONDREJKA, 2004b).

By default, objects are created in SL as non-physical entities, i.e., with the *'Physical'* attribute disabled (PHYSICAL, 2007). An object may be made physical by setting the *'Physical'* checkbox in the edit-object dialog box or using the LSL functions *llSetStatus* or *llSetPrimitiveParams*.

Physical objects can, in principle, be affected by 'wind', 'gravity', and collisions with other objects (NON-PHYSICAL, 2008). 'Gravity' is realized through an invariable force applied to every physical object of its mass times 9.8 m/s$^2$ in the negative *z* direction (GRAVITY, 2006). However, physical objects will not keep an exact and constant 9.8 m/s$^2$ gravity acceleration during a free fall because some wind resistance seems to act to allow falling objects to achieve terminal velocity, as it was observed by dos Santos (2009). 'Wind' affects avatar clothing/hair and trees[5] and can affect particles[6] and flexible prims[7]. It 'naturally' (programmatically) varies in velocity and direction, based on the 2D, stable fluid method, with some pseudo-stable chaos effect added (llWind, 2009). Wind does not cause friction (or *drag forces* (DOS SANTOS, 2009)), though, and no air or water resistance was implemented as it is numerically unstable, especially for small objects, and requires a lot of tuning to the physics engine and parameter space to make it work right (llWind, 2009). Therefore, physical objects are, in practice, subject only to 'gravity' and collisions. Notice that as there is no "real" air resistance in SL, any impulse off the vertical (gravity) axis will cause the object to keep moving forever (llApplyImpulse, 2009).

As seen before, a few SLS functions, called *kinematic functions*, such as *llSetPos* and *llSetRot*, do not work for physical objects. Other functions, called *kinetic functions*, like *llSetForce* and *llSetTorque*, operate on physical objects only. There are also functions that work on both physical and non-physical. In addition, some other functions such as

---

[4] http://wiki.secondlife.com/wiki/LSL_Portal.
[5] Linden trees are decorative detailed one-prim objects that can be placed on the ground or underwater. They will bend in the local simulated wind, but do not otherwise do anything interesting (TREE, 2006).
[6] Particles are entirely client-side generated, free-floating, non-object (sprites) visual effects that can simulate blood, dust, explosions, fire/flames, gas, smoke, sparks, spray, steam, waterfalls, waves, weather, etc. (PARTICLES, 2010).
[7] Flexible prims, also called "flex prims" are prims apparently made flexible through a client-side effect (FLEXIBLE, 2006) used to simulate flowers, curtains, etc.

*llRezObject* (llRezObject, 2009) and *llTargetOmega* (llTargetOmega, 2010) do not work as expected for non-physical objects:

*llRezObject:* non-physical objects are rezzed but stationary (llRezObject, 2009);

*llTargetOmega:* non-physical objects are only apparently rotated on the client ('client-side') while physical objects are rotated on the server ('server-side') (llTargetOmega, 2010).

It should be noted, however, that the motion resulting from a kinematic function is determined by the mass (inertia) of the object, which depends only on its size and shape, not on its material (MATERIAL, 2012). To complicate things further, as discussed in dos Santos (2009), the effect of a kinetic function will depend on the 'energy' content of the object which is continuously spent while the function acts reducing its performance.

*There are no fluids in SL*

There are no fluids in SL. In SL 'Oceans', 'Water' is a property of Regions and can be covered with soil but cannot be contained and has no effect on avatars or objects (WATER, s.d.). Outside this situation, it is a mere texture applicable to solids, even when making a swimming pool. Fountains are made of a different class of objects named *particles*[8]. It can, however, be made further 'realistic' by animating the texture with scripts (WEBER et al., 2007, p. 258).

*SL Physics is restricted to Mechanics*

The physics engine Havok deals with game-genre specific problems like vehicle simulation, human ragdolls, physical interaction of keyframed characters within a game environment, and character control (HAVOK.COM, 2008, p. 96). As discussed by dos Santos (2009), any possible electromagnetic or nuclear interaction simulations is precluded because Havok does not simulate any physics beyond Mechanics.

*SL Time is not consistent*

As discussed by dos Santos (2009), time in SL can be affected by script generated simulator lag, network lag, and server load, and, therefore, may not be particularly accurate (llGetRegionTimeDilation, 2009). While SL is able to run qualitative experiments and to cope with simple mechanics experiments with a corresponding decrease in accuracy, it will not give response time down to milliseconds consistently (SECOND Life: Guidelines for Educators, 2008, Technical essentials, § 5).

*SL Mass depends only on the size of the object*

Contrary to the Newtonian definition of mass, SL object mass depends only on its size and shape, not on its material type, (MASS, 2009). As a consequence, avatar mass depends only on its height and attachments will not affect avatar mass, except for shoes, which change avatar height and, therefore, its mass (DOS SANTOS, 2009). Notice, however, that there is a parameter called GRAVITY MULTIPLIER (GRAVITY Multiplier, 2011) in the *llSetPrimitiveParams* LSL function that when increased has a similar effect as increasing the object's mass.

*SL Energy controls how effectively scripts act on objects*

---

[8] *See supra* note .

As discussed by dos Santos (2009), differently from its definition in Physics, energy in SL is a dimensionless quantity implemented to limit the amount scripts can use a number of dynamics functions. An object expends energy when functions try to change its motion, while objects continuously receive energy from the SL grid at a rate of 200/mass units of energy per second until the 100% full energy limit (DOS SANTOS, 2009). Kinetic functions demand energy at different rates and some of them may even not be able to act on heavy objects if they reduce object's energy faster than the grid can refill it (DOS SANTOS, 2009).

*SL Buoyancy does not take water level into account*

In Physics, buoyancy is the upward force that a fluid exerts on an object that is less dense than it is. Buoyancy allows a boat to float on water and provides lift for balloons. However, as water has limited meaning in SL, buoyancy does not take water level into account: the object will float up the same rate whether it is under or above water (llSetBuoyancy, 2009). As discussed by dos Santos (2009), the buoyancy of an object in SL is a dimensionless quantity such that values between 0.0 and 1.0 result in a gentler than regular fall. Setting it to exactly 1.0 will cause the object to float as if no gravity existed and a buoyancy value greater than 1.0 will make it rise, what is useful to build balloons. Negative buoyancy values are accepted and will simulate an increased downward gravitational force.

*SL Light do not propagate*

The main illumination in SL comes from the Sun and the Moon replicas, being six additional sources allowed for incidental lighting. However, the graphics cards in most home computers prevent perfect shadows from being generated in real time, and opaque obstacles are ignored. Therefore, objects do not cast shadows while light cannot be contained by walls of a room (WEBER et al., 2007, p. 33). As a consequence, as discussed by dos Santos (2009), light merely "is there" in SL, without any physical mechanism involved in its production or propagation, as in the most primitive conceptions of light.

*Representation of physical quantities in real time*

Physical quantities are not automatically represented by the SL client as it happens in *Modellus*. Nevertheless, LSL provides functions to access directly some of them, e.g. *llGetPos* and *llGetVel* that return the position and velocity of the object in the region, respectively. Therefore, these quantities can be represented directly, in real time, during the entire duration of the phenomenon as shown by dos Santos (2012). There are no functions in LSL to obtain the value of some other relevant quantities, such as kinetic energy, momentum, in the correct scalar or vector form, which can, however, be easily calculated by the script from those fundamental quantities. In addition, dos Santos has already demonstrated how scripts can make use of the SL resources for vectorial representations (2012).

*SL Physics is not Newtonian Physics*

Second Life has been quite mistakenly labeled as 'a three-dimensional online computer simulation of the *real world* [italics added]' (LINDEN Vehicle Tutorial, 2012). It is worth remembering, however, that 'Physics in Second Life is not real world physics' (LINDEN Vehicle Tutorial, 2012). According to Philip Rosedale, founder of Linden Lab, the intention behind SL was to conceive "a world in which everything was built by […] the people who were there in a kind of Lego block sort of way *to rebuild the laws of physics* [italics added]" (ROSEDALE, 2007b). As a matter of fact, it was already shown by dos Santos (2009) that SL Physics is neither the Galilean/Newtonian "idealized" Physics nor a real-world Physics virtualization. Rather, that author concluded that it is hyper-real and that it, nevertheless, provides resources for building *surreal* Physics microworlds where physical laws are different or changeable by the student – a 'surreal' microworld in some sense, rescuing Papert's (1980) never-implemented proposal.

*SL is not a usual simulator*

In comparison with a traditional simulator such as *Modellus* (TEODORO et al., 1997), SL lacks some fundamental functions to provide initial conditions to objects that simulators usually offer. There is nothing like *llSetVel* to set the initial velocity of the object, and the existing *llSetPos* function does not act on physical objects (llSetPos, 2011). As Andrew Linden, Co-Founder of Linden Lab, commented once (2010), moving an object at a constant velocity is very hard at present SL development status. In general, the only option to set an object into motion are the functions *llSetForce* and *llApplyImpulse*; but one should notice that they will not deliver steady velocities – quite on the contrary, the object will start to gain speed (to accelerate) in the first case and to lose speed in the second, due to damping (more correctly, to drag). However, one has to concede that we do not have such resources in the real world either – we can only accelerate (by pushing or hitting) and slow down objects. In this sense, SL Physics is again more real than classroom Physics (DOS SANTOS, 2009).

This difficulty can be seen in the three concrete examples discussed by dos Santos (2012):

- A replica of an *air track* in which the builder found herself unable to have the glider to move along a track at a given constant velocity even if using buoyancy to cut down on friction.
- A simulator themed as a pair of amusement park non-wheeled bumper cars running on rail tracks and performing head-on collisions in which the author had the same sort of troubles in controlling the velocity in SL as in the previous example
- A 'cannon' that can simulate both Newtonian Mechanics and Buridan's Impetus Theory as per user's choice where the author found it tricky to counterbalance gravity and to achieve a rectilinear trajectory simulation for the cannonballs.

We now proceed to analyze the SL environment according to Narayanasamy et al. (2006) and to Johnston and Whitehead (2009) criteria to determine into which of *training simulators, games*, *simulation games,* or *serious games* categories SL fits best.

## *ANALYSIS*

Here, we will analyze SL by each one of the chosen criteria as described in the Methodology section.

*Analysis according to Narayanasamy et al. (2006)*

These authors consider that the distinction between Games, Simulation Games, and Simulators lies in the fact that games are designed to be "entertaining, fun, and engaging," while training simulators are designed to "qualify and track the development of specific skills in its operators". With this principle in mind, they have built a common taxonomy for these systems that is summarized in Table 1.

*Table 1 - Identifying Games, Simulation Games, and Simulators*
*(© 2006 Narayanasamy et al. Used with permission.)*

|   | **Identifying Characteristics** | **Games** | **Simulation Games** | **Training Simulators** |
|---|---|---|---|---|
| 1 | Involves simulation | 1. A virtual environment is present.<br>2. The application interactively engages the user in a form of simulation. | | |
| 2 | Imaginative experience | 1. May provide an imaginative or fictitious simulated environment. | | 1. Provides only recreations of real world environments. |
| 3 | Entertaining, Fun & Engaging | 1. Provides entertainment.<br>2. Provides interesting & engaging challenges.<br>3. Provides a fun experience. | | 1. Not intended to be entertaining, fun, or engaging.<br>2. Operator can possibly find the application entertaining, fun, and engaging. |
| 4 | Skills development | 1. Does not provide an application specific skill development.<br>2. Possible, although not as a primary feature. | | 1. Operator skills development is the primary purpose of a Simulator. |
| 5 | Type of Challenge | 1. Ideally a continuous and intelligent challenge. | | 1. Challenges depicted accurately with respect to an equivalent real world scenario. |
| 6 | Gestalt | 1. Presence of game-play patterns.<br>2. Game play patterns may vary.<br>3. Possible development of a game play gestalt. | | 1. Presence of standard operational procedures.<br>2. Procedures do not change. |
| 7 | Goal–Oriented | 1. Goal-Oriented Activity present. | | 1. Goal-Oriented Activity absent.<br>2. No obvious end state. |
|   |  | 2. End State Present. | 2. No obvious end state. |  |

Following Narayanasamy et al. (2006), we now examine each identifying characteristic in Table 1 carefully, trying to decide for each characteristic which category SL belongs to.

**1. Involves simulation**. SL does provide a virtual environment where users interactively engage in a form of Life/Social simulation of an earthlike world. Not the real world, however, but rather an imaginative and fictitious simulated world in which an avatar can fly, teleport, change its appearance from a witch to a dragon, to a pillar box or to anything else it wishes. A world which dos Santos (2009) labeled 'hyper-real' as it follows neither the real world physical laws nor the Galilean/Newtonian ones but "offers a set of capabilities, which are in many different ways superior to the real world" (ROSEDALE, 2007a).

**2. Imaginative experience**. The rich and massively multi-user 3D simulated experience it provides is sensationally pleasurable, engaging the user to explore the territory and uncover or discover new previously hidden functions, like flying, changing appearance, creating objects, buying accessories, spending *Linden Dollars* (the virtual currency in SL), and so on. It is a fictitious experience that reach into the pupil's imaginations (DE FREITAS; GRIFFITHS, 2009, p. 57).

**3. Entertaining, Fun & Engaging**. While SL was intended to be entertaining, fun, and engaging, it may posit the above-mentioned non-explicit challenges to *newbies*, who would have to overcome them to experience SL as such. This, in fact, could account for the reduced users' retention. In any case, users are free to express themselves even if less socially accepted behaviors such as walking naked, having sex, and shooting are restricted to 'adult' or 'game' areas. According to Ondrejka (2004b), SL's design was focused on fostering creativity and self-expression in order to create a vibrant and dynamic world full of interesting content.

**4. Skills development**. SL is clearly not focused on operator skills development besides those needed to make the avatar walk, fly, chat, and so forth.

**5. Type of Challenge**. There are no clearly depicted challenges besides those related to making the avatar walk, fly, chat, and so forth.

**6. Gestalt**. Narayanasamy et al. (2006) make use of (LINDLEY, 2003) concept of *gameplay gestalt*, understood as a pattern of interaction with the game system, as allowed by its rules, learning to play it, in a way that supports progress, with persistence and basic ability, eventually completing or winning a game. However, we found no evidence of gameplay gestalt developments besides those related to the above-mentioned non-explicit challenges. Besides, there are no standard, invariable operational procedures. Residents of SL participate in a wide variety of activities, from building homes and holding parties to entrepreneurial activities of all kinds (ONDREJKA, 2004b). Therefore, SL disqualifies itself from being labeled as a training simulator well.

**7. Goal-Oriented**. There are no goal-oriented activities, competition, conflict, or rules imposed on the residents beyond reasonable restrictions on improper behavior and the physical rules that guarantee its physical verisimilitude to the real world. There is no obvious end state as well.

From the discussion above, one sees that SL fails to fit itself clearly in any of the Games, Simulation Games, or Simulators categories. Our conclusion, agreeing with Bestebreurtje (2007), is that, by Narayanasamy et al. (2006) criteria, *SL is neither a game nor a training simulator nor even a simulation game*.

*Analysis according to Johnston and Whitehead (2009)*

We now proceed to analyze SL according to the other chosen criteria.

Johnston and Whitehead (2009) start by disagreeing with the work of Narayanasamy et al. (2006) and proposing that the distinction between games and simulation games from simulators "be made on the basis of intent and closeness to the user's reality, as opposed to designer's intent", as was the case of the previous authors.

Johnston and Whitehead (2009) define a game, both traditional games and video games, as a closed formal system that represents a subset of reality. Then they define serious games, as a subset of games used for explicit learning purposes and training simulations as a subset of serious game that closely resemble the user's own specific reality. They stress that their meaning of "closeness to reality" is more related to the skills and processes required of the user of a training simulation than the closeness of resemblance to the reality of its visual appearance.

For these authors, when the primary goal of the game is education, it may be classified as a serious game. If the serious game closely resembles the player's reality, it is then a training simulation.

Narayanasamy et al. (2006) proposes that one of the main distinguishing characteristics between games and simulators is the presence of goals in the first and of objectives in the second. Johnston and Whitehead (2009), on the other hand, argue that both goals and objectives are present in both games and simulators, that the goals may depend on the player's definite intentions and that broad based goals rather than restricted objectives are found in many simulations.

From the discussion about particular features of the SL in the previous section, we have to concede that it is a closed formal system that represents a subset of reality and that, therefore, depending of its user intents, it could have to be classified as a game by Johnston and Whitehead (2009) criteria. As a matter of fact, one can find thousands of entries within the Yahoo!®Answers site asking about how to play *Second Life game*, even if Philip Rosedale himself vehemently denied it once: *"I'm not a gamer, and SL isn't a game. From the start, we/LL […] focused on making SL very exciting and visceral and inspirational, but not on making it a game."* (ROSEDALE, 2006a)

Of course, each SL content creator has her own goals and, therefore, anything created within SL with educational oriented goals would be classified as a serious game by these authors' criteria while nothing else would be. Depending on the closeness of the resemblance to the reality in terms of the skills and processes required of the user it can be further be classified as a training simulator. If we agree to abide by these criteria, the air track replica discussed on (DOS SANTOS, 2012), due to its resemblance to the real-life laboratory counterpart, would be categorized as a training simulator while the Brownian motion simulation and the bumper cars head-on collision simulator also discussed on (DOS SANTOS, 2012) would be as serious games, all three being operated inside the serious game SL itself.

Maybe Azzara (2007) was the one to come closest to a definition for SL: "*Linden Labs had the brilliantly insightful idea to put out into the world what amounts to a multiplayer video game platform with no game!*" Legend says that, during a 2001 meeting with investors, Rosedale noticed that the participants were more responsive to the collaborative, creative potential of SL than to his haptics development for which the virtual world was built. As a result, the initial objective-driven, gaming focus of SL was accordingly shifted to a more user-created, community-driven experience (ROSEDALE, 2006b; WHITSON, 2010). Castronova (2004) argues that virtual worlds are manifesting themselves with two faces: one invoking fantasy and play, the other merely extending day-to-day existence into a more entertaining circumstance, the latter aspect beginning to dominate the former, gradually blurring this distinction. If this is true, this discussion about SL being a game or not will soon become quite immaterial. Meanwhile, we may consider SL as a *game platform with no game*, an enormous and sophisticated simulator of an entire Earthlike world used to simulate real life in some sense and a viable and flexible state-of-the-art physics engine-powered platform for microworlds and simulations.

Unfortunately, however, one cannot say that SL is an easy-to-use platform. Most users agree that a high learning curve exists for new users (SANCHEZ, 2009), which means that any proposal of using SL for teaching should reserve several hours, just to have the students become familiar with essential tasks, e.g. walking, pass through doorways, go up stairs, manipulate objects, and so forth.

We consider that *Modellus* (TEODORO et al., 1997) main innovation was to provide users with a very accessible interface that eliminated the need to program graphic interfaces in personal computer working in DOS. Our vision is that SL is equally innovative as a 3D platform that could be made much more accessible to novice users with a more user-friendly interface, perhaps without forcing them to enter into the depths of LSL programming, reducing the above-mentioned learning curve and making such an interesting tool as SL available to a number of teachers.

## CONCLUSION

As stated before, the main objective of this study was to examine the viability of SL as an environment for physical simulations. We consider that the rich immersive 3D massively multi-user experience SL provides is pleasant, engages the user to explore the territory, and, therefore, offers a number of advantages over a 'traditional' 2D simulator. Sanchez (2009) affirms that "designers can create a user experience that will build on the strengths of the virtual world while overcoming the obstacles". We agree with Sanchez and consider that, despite the restrictions and differences in comparison to a 'traditional' simulator such as Modellus, SL shows itself as a viable and flexible state-of-the-art physics engine-powered platform for microworlds and simulations, even if it requires some creativity to overcome the difficulties of implementation.